\documentclass[prc,preprint,showpacs,superscriptaddress,amsmath]{revtex4}

\usepackage{graphicx}
\usepackage{dcolumn}
\usepackage{graphics}
\usepackage[dvips,usenames]{color}

\newcommand{\vs}{\quad}

\newcommand{\pv}{\mbox{\boldmath $p$}}
\newcommand{\av}{\mbox{\boldmath $a$}}
\newcommand{\xiv}{\mbox{\boldmath $\xi$}}
\newcommand{\Jv}{\mbox{\boldmath $J$}}
\newcommand{\vv}{\mbox{\boldmath $v$}}
\newcommand{\nabv}{\mbox{\boldmath $  \nabla$}}
\newcommand{\dt}{{\mathrm d}}
\newcommand{\tr}{{\mathrm t}}
\newcommand{\slp}{{\mathrm{sl}}}
 
\newcommand{\be}{\begin{equation}} 
\newcommand{\ee}{\end{equation}} 
\newcommand{\ba}{\begin{eqnarray}} 
\newcommand{\ea}{\end{eqnarray}} 
\newcommand{\bse}{\begin{subequations}} 
\newcommand{\ese}{\end{subequations}}

\begin{document}


\title{Non-Gibbs Particle Spectra from Thermal Equilibrium}
\author{Tam\'as S. Bir\'o} \affiliation{ 
  University of Giessen,
  D-35394 Giessen, Heinrich-Buff-Ring 16, Germany  
}
\affiliation{  
 KFKI Research
  Institute for Particle and Nuclear Physics, H-1525 Budapest P.O.Box
  49, Hungary} \quad  

\author{G\'eza Gy\"orgyi} \affiliation{Institute for
  Theoretical Physics, HAS Research Group, E\"otv\"os University,
  1117 Budapest, P{\'a}zm{\'a}ny s{\'e}t{\'a}ny 1/a, Hungary} 

\author{Antal Jakov\'ac} \affiliation{Research
  Group for Theoretical Condensed Matter of HAS and TU Budapest,
  H-1521 Budapest, Hungary}

\author{G\'abor Purcsel} \affiliation{
  University of Giessen,
  D-35394 Giessen, Heinrich-Buff-Ring 16, Germany  
}
\affiliation{
  KFKI Research
  Institute for Particle and Nuclear Physics, H-1525 Budapest P.O.Box
  49, Hungary}  

\date{printed \today}

\begin{abstract}
  We propose that transverse momentum spectra with power-law tails
  observed in ultrarelativistic heavy ion collisions can be interpreted as
  originating in a medium in thermal equilibrium.  General conditions
  on the dynamical equations are formulated, leading in equilibrium to
  various energy distributions of particles.  Starting with the linear
  Fokker-Planck equation we analyze conditions for Boltzmann-Gibbs,
  Tsallis, or other equilibrium distributions based upon the
  dependence of fluctuation and dissipation on the energy of the
  observed subsystem.  RHIC neutral pion data definitely exclude
  Boltzmann-Gibbs distribution over the entire range of observed
  transverse momenta.
\end{abstract}
\maketitle


\vs
\section{Introduction}

There is an increasing interest in phenomena, where a non-conventional
equilibrium distribution, different from the Boltzmann-Gibbs (BG)
formula \cite{Gibbs,Balescu}, arises.  Like the BG weight,
$\exp(-E/T)$, can be derived as the canonical probability starting
from Boltzmann's entropy formula, generalized entropies may lead to
other canonical distributions.  Frequently discussed suggestions are
the R\'enyi \cite{RenyiBook, Wehrl78}, or the Dar\'oczy-Tsallis
entropy \cite{Daroczy,AczelDaroczy,Tsallis,PratoTsallis,LRT01,LRT02,
  TRLB,TsinNonextensiveEntropy}, giving rise to non-extensive
thermodynamics.  The corresponding canonical distribution shows a
power-law tail at high energies, which can be observed in several
physical systems.  Among them the transverse momentum spectra of high
energy particles measured in electron-positron, proton-proton and
heavy ion collisions are particularly interesting: due to the high
event statistics relative yields down to $10^{-9}$ can be
experimentally measured \cite{PTLexperiments}.  Since the BG
distribution in equilibrium is ubiquitous in atomic, molecular or
solid state physics, the high precision knowledge pointing towards
deviations from BG, obtained from elementary particle and heavy ion
physics, may help us to gain new insights into statistical physics,
too.

\vs The BG case has ever been surmised on the ground of being the
simplest set of assumptions which can be made about a statistical
system in physics. The Tsallis distribution is already a
generalization, leading back to the BG case at a certain value of a
further parameter besides temperature, the Tsallis index. Establishing
some unconventional statistics, however, is never an easy enterprise:
Microdynamical calculations with many fine details should lead to the
assumed canonical distribution compatible with the assumed formula for
the generalized entropy in one or in the other case.  Models of
diffusion, solving a stochastic Langevin equation and statistically
analyzing these solutions with the help of a Fokker-Planck equation,
can be viewed as a step towards a microscopic approach to the BG
distribution.  For the Tsallis type generalization of the exponential
distribution several model suggestions occurred already, which seek a
dynamical mechanism for generating power-law tailed distributions in
particle physics \cite{Beck00}.  Works discussing a fluctuating
temperature \cite{Wilk00,Wilk02} or an energy imbalance due to medium
effects in two-body collisions \cite{ShermanRafelski} successfully
convert the assumption of Tsallis distribution of energy into an
assumption of an underlying gamma distribution of the inverse
temperature.  Some generalizations of the Fokker-Planck equation to a
nonlinear version explain this step by assuming diffusion in a phase
space with fractional dimension
\cite{LisaBorland98, Zanette99, Kaniadakis01}.  An assumption of a
stochastic component in the damping constant in the original Langevin
equation (which also leads to a Tsallis distribution in equilibrium)
\cite{Biro+Jako} connects this non-extensive thermodynamics with the
noisy behavior of nonabelian plasmas
\cite{Arnold99, Arnold00, Bodeker99, Bodeker01, Bodeker02, Muller04}.
 
\vs Effective stochastic equations have also been derived in the
framework of field theory. For one of the simplest and most general
field theories, the $\Phi^4$ model, dissipative and stochastic
contributions to the evolution of long-wavelength modes have been
obtained by using the Feynman-Vernon influence functional approach
\cite{Greiner98, GreinerThesis, Greiner04}.  The general form of this
effective equation of motion in a Markovian approximation (eq.\ (4.66)
of Ref.\ \cite{GreinerThesis} contain additive and multiplicative
noise terms. They can, in principle, be combined into one physical
noise, $\xi = \xi_1 + \Phi \xi_2 + \Phi^2 \xi_3$, whose
auto-correlation depend on the field itself, and therefore on the
energy of the subsystem represented by this field modes, too.  In
\cite{RafelskiWalton} the equilibration of charm quarks in hot quark
gluon plasma has been shown by Walton and Rafelski to lead to a nearly
Tsallis distribution (cf. Fig.~1 of \cite{RafelskiWalton}), based on
perturbative QCD estimates of damping and stochastic forces. 

\vs The question arises, whether some general rules can be formulated
that connect the stochastic microdynamics with a generalized entropy
formula, the latter serving as a basis for the thermodynamical
treatment. Quite a few steps in this direction have been already done.
As we mentioned it already, generalizations of the Fokker-Planck or
Langevin equation approach include multiplicative noise
\cite{Biro+Jako}, non-linear dependence on the particle- or
probability-density \cite{LisaBorland98, Zanette99, Kaniadakis01}, or
a generalized linear response theory of thermodynamical forces
\cite{Frank99}.

\vs Since at low enough energy the experimental Tsallis distribution
is practically indistinguishable from the BG distribution, improved
analyzes of experimental data on particle distributions are desired in
order to decide whether one or another generalized statistics occurs
in nature (and with what parameters).  In analyzing particle spectra
one has to relay on some assumptions about the state of the system
emitting the particles.  Our main proposition in this paper is that it
is possible to interpret the data on the basis of local thermal
equilibrium of the quark-gluon system.  We do not exclude that
non-equilibrium effects can play a role, rather we demonstrate a way
that can lead to non-BG equilibrium distribution.  In other words, we
propose that deviation from BG statistics does not imply that the
medium cannot be in equilibrium.  Furthermore, one should be aware
that through the spectra one studies the energy distribution of a
subsystem, i.e. the individual hadrons created in the collision.  In
the case of strong coupling, or strong correlations, however, the
distribution of a subsystem can assume many forms, for example, it is
not necessarily BG even if the whole system was BG. A remarkable
example of such a situation is the Tsallis type distribution obtained
in a BG environment for charmed quarks \cite{RafelskiWalton}.
Nevertheless, one is tempted to believe that emitted particles from a
system in thermal equilibrium, in the process of nearly instantaneous
hadronization, reflect the equilibrium distribution therein.  We
shall, therefore, work under the assumption that the spectra are
faithful images of the equilibrium statistics around the instant of
the creation of the hadrons.

\vs
\section{Stochastic dynamics} 
 
\subsection{Energy dependent noise in one dimension} 
 
\vs In this section we present a simple, Markovian model of linear
kinetics for one phase variable, the momentum, in order to arrive at
non-BG distributions as stationary solution.  Our main proposition is
that the energy, $E(p)$, representing an arbitrary dispersion relation
for the quasiparticle plasmon, enters the Langevin equation, which may
then give rise to quite a wide variety of equilibrium distribution
functions of $E(p)$.  We take the damping term proportional to the
group velocity of the plasmon in the medium, $v_g=E^\prime(p)$, and
include a general white noise term, $\xi$, whose autocorrelation
contains an energy dependent diffusion coefficient.  Then we wind up
with the Langevin equation

\be \dot{p} + G(E) E^\prime = \xi, 
\label{LANGEVIN}  
\ee  

where the noise obeys 

\be    
   \langle \xi(t) \rangle = 0, \qquad 
   \langle \xi(t)\xi(t') \rangle = 2 D(E) \delta(t-t'). 
\label{NOISE} 
\ee

Note that in the drift term $G(E(p))$ and the diffusion coefficient
$D(E(p))$ we have here a particular case of the more general $p$
dependence.  We chose this to demonstrate how one can build up a
dynamical treatment leading to a stationary distribution that depends
only on the energy $E$. In the general situation, besides the $E(p)$
dependence, $p$ can also appear explicitly, implying some modification
in the Langevin equation, but for the sake of simplicity we omit this
from the present demonstration.

\vs In the case of a non-constant diffusion coefficient one should
specify the type of discretization in time of the Langevin equation
\cite{Stratonovich}.  We shall adopt the Ito convention for purely
practical reasons: then the corresponding Fokker-Planck equation has
the simplest form.  This does not imply any physical restriction,
because a Langevin equation defined with arbitrary discretization can
be transformed into another one with the Ito rule.  Then eq.\ 
(\ref{LANGEVIN}) leads to the Fokker-Planck equation, which can be
written as a continuity relation 

\bse
\label{FOKKER-PLANCK}
\ba
0 &=& \frac{\partial f}{\partial t} +\frac{\partial J }{\partial p},
\label{eq:continuity} \\ 
J &=& - G(E) E^\prime f \, - \, \frac{\partial D(E) f}{\partial
  p},
\label{current}
\ea \ese 

where $f(p,t)$ is the time-dependent distribution for the momentum.
The stationarity condition is obviously that $J$ should be constant,
while a vanishing $J$ can be interpreted as the stricter condition of
detailed balance \cite{vanKampen}.  Even the latter requirement is met
if the stationary distribution is

\be f_s(p) = f(E(p)) = \frac{A}{D(E(p))} \exp \left(-
  \int_{E_0}^{E(p)} \frac{G(\bar{E})d\bar{E}}{D(\bar{E})} \right).
\label{STAT-DISTR}
\ee 

We assume that $f_s(p)$ is normalizable, then $A$ is the normalization
factor.  Specifically, it is determined by the plasmon density
integral for one particle in a volume $V$, that is, $\int f(E(p)) \dt p =
1/V$.  If we exponentiate also the diffusion coefficient from the
prefactor then the so created exponent is traditionally denoted by
$-U(p)$, where $U(p)$ is called stochastic potential \cite{Gardiner}.
The stationary distribution is thereby

\be 
f_s(p) \, = \, A \, \exp\left( - U(p)\right),
\label{FORMAL-GIBBS}
\ee

a formula reminiscent to the BG distribution when $U$ is identified
with the energy per temperature, playing a central role in kinetic
theories of thermalization, and having numerous physical applications
(see \cite{Graham89} and refs.\ therein).  In our present case,
however, we have already a ``static'' energy $E(p)$ assumed to be
known prior to and independently from the stochastic dynamical
considerations, e.g.\ it is a quasiparticle dispersion relation in an
interacting medium.  This is an important point of our theory, since
in terms of the physical energy $E$ the stationary distribution
(\ref{STAT-DISTR}) is arbitrary to a large extent.  So the present
framework of a Fokker-Planck equation linear in the probability
distribution can still give rise to non-BG stationary statistics in
terms of $E$.  Naturally, the BG-like formula (\ref{FORMAL-GIBBS})
remains valid for the same statistics, but if $U$ is not the physical
energy, one no longer has BG thermodynamics.

\vs It is useful to introduce the energy dependent inverse logarithmic
slope of this distribution as

\be T_{\slp}(E) = - \frac{1}{(\ln f(E))^\prime},
\label{INV-LOG-SLOPE}
\ee

where $~^\prime$ means derivative in terms of the energy and the
subscript stands for ``slope''.  One immediately sees that the
stationary distribution (\ref{STAT-DISTR}) can be given in a form,
which appears as a natural generalization of the BG formula, as

\be f(E) \: = \: A \, \exp\left(- \int^E \:
  \frac{d\bar{E}}{T(\bar{E})} \right).  
\ee 

In this sense $T_{\slp}(E)$, the energy-dependent inverse logarithmic slope
of the energy spectrum, is a generalization of the temperature for
non-BG distributions.  We recover the traditional BG distribution if
and only if $T_{\slp}(E)$ does not depend on the energy, and then
$T_{\slp}(E)\equiv T$, the BG temperature.

\vs From the stationary distribution (\ref{STAT-DISTR}) we get

\be
G(E) = D(E)/ T_{\slp}(E) - D'(E),
\label{eq:FDT-1d}
\ee

a relation which amounts to a generalized fluctuation-dissipation
theorem.  Indeed, given a stationary distribution $f(E)$ this is the
condition the damping and diffusion coefficients must satisfy so as to
be compatible with the prescribed distribution.  If we presume the
damping coefficient $G(E)$ to be known, (\ref{eq:FDT-1d}) can be
solved for the diffusion factor as

\be {D}(E) = \frac{1}{f(E)} \int_E^{\infty} {
  G}(\bar{E}) f(\bar{E}) d\bar{E},
\label{D-FROM-G-1d}
\ee 

where the integration constant is set by our assuming that there is no
upper cutoff energy.

\vs For illustration we discuss the simple case of a constant damping
$G=G_0$.  Even then the diffusion coefficient is not necessarily
constant in $E$, unless the stationary distribution $f(E)$ is the BG
distribution.  In the latter case we get, by substituting $f(E)=A
e^{-E/T}$ into (\ref{D-FROM-G-1d}), the known Einstein relation, where
both damping and diffusion factors are independent of $E$, so $D=D_0$
where 
  
\be {D_0} = T {G_0} \label{eq:simple-FDT-1d}.  \ee 
  
This is the simplest form of the fluctuation-dissipation theorem.
However, we now see that the widely practiced reference to the
fluctuation-dissipation theorem only, when assuming a proportionality
like in (\ref{eq:simple-FDT-1d}) is, in principle, not sufficient,
since tacitly it is also understood that $G$ and $D$ are independent
of $E$, quite a particular case.  When the $G$ and $D$ may be
$E$-dependent, then one should resort to the fluctuation-dissipation
theorem in the form (\ref{eq:FDT-1d}).  However, in ranges of near
constancy of $G$, approximate proportionality like
(\ref{eq:simple-FDT-1d}) can hold.
 
\vs Considering non-BG distributions, still in the particular case
when $G=G_0$ is constant, we can distill eq.\ (\ref{D-FROM-G-1d}) into
   
\bse
\label{Einstein-all} \ba {D}(E) &=& T_{\mathrm E}(E) {G_0},
\label{EinsteinRelation} \\
T_{\mathrm E}(E) &=& - \frac{1}{\left(\ln \int_E^\infty f(\bar{E})
    d\bar{E}\right)^\prime}.
\label{EinsteinTemperature}
\ea \ese 

This can be considered as the generalization of Einstein's relation
between viscosity and diffusion coefficient for an arbitrary
stationary distribution, where the Einstein temperature $T_{\mathrm
  E}(E)$ is the proportionality factor.  The latter is generically
different from the inverse local slope function $T_{\slp}(E)$ of
(\ref{INV-LOG-SLOPE}), apart from the BG case, when $T_{\mathrm E}(E)=
T_{\slp}(E)=T$ constant.  So we see an interesting specialty of BG
thermodynamics: the temperatures defined by the inverse logarithmic
slope of the distribution and the ratio of the diffusion and viscosity
coefficients coincide, while for any other distributions they are not
the same temperature functions.  We stress that the Einstein
temperature (\ref{EinsteinTemperature}) enters the
fluctuation-dissipation relation as a proportionality factor only in
the case of constant mobility ${G(E)=G_0}$, else we do not have a
physical interpretation for it presently.

\vs In what follows we demonstrate that the Tsallis distribution
arises as, in a sense, the simplest extension beyond the BG statistics.
The generalized temperature is (i) in the simplest case
$T_{\slp}(E)\equiv T$ constant, this corresponds to the BG statistics,
and (ii) in the next-to-simplest case $T_{\slp}(E)$ is linear in $E$.
The most elementary realizations of them with $G(E)=G_0$ are (i)
$D(E)=D_0$ constant, and when (ii) $D(E)=D_0+D_1 E$ is linear.  The
former case, (i), leads to the BG distribution
  
\be f_0 = A \exp(-E/T),
\label{GIBBS} 
\ee 
 
with $T=D_0/G_0$ being the temperature. The diffusion constant $D_0$,
the damping constant for a particle with mass $m$, $\gamma=G_0/m$, and
the free particle dispersion relation $E=p^2/2m$ lead to the
well-known description of Brownian motion.  The linear case (ii)
yields a power-law tailed distribution, often cited as the Tsallis
function
  
\be 
f_1 = A \left(1 + (q-1)\frac{E}{T} \right)^{-q/(q-1)},
\label{TSALLIS}
\ee 
  
with the Tsallis temperature parameter $T=G_0/D_0$, the same formula
as in the case (i), and the so called Tsallis index $q=1+D_1/G_0$.
The temperature function (\ref{INV-LOG-SLOPE}) now is 
 
\be 
T_{\slp}(E) = \frac{T}{q} + \left(1-\frac{1}{q}\right) E,
\label{TSALLIS-TE}
\ee 
 
indeed linear in $E$.  Concerning the Einstein temperature
(\ref{EinsteinTemperature}), it coincides with $T$ in the BG case (i),
and in the Tsallis case (ii) we obtain $T_{\mathrm E}(E)=qT_{\slp}(E)$
with (\ref{TSALLIS-TE}), so $T_{\mathrm E}(E)$ is also linear in $E$.
In sum, we have BG if the dynamical coefficients and the temperature
are energy-independent, while, for constant damping, the Tsallis
statistics corresponds to generically linear $E$-dependence of various
physical quantities.  So from the viewpoint of energy-dependence, it
is justified to call the Tsallis ensemble the simplest next to the
canonical BG case.  It is also apparent from the above reasoning, that
based on dynamical considerations statistics more complicated than
Tsallis are in principle allowed.

\vs Besides the BG case (i) also in (ii) a thermodynamical treatment
in equilibrium has been widely studied in the last decade, inspired by
the seminal works \cite{Daroczy,Tsallis}.  In this section we
complemented the static picture in revealing some non-equilibrium
properties of systems, describable by linear evolution equation for
distributions, and relaxing to non-BG, in particular to Tsallis,
ensembles.


\vs
\subsection{Phase space of arbitrary dimensions}

\vs The generalization of the above discussion for state variables in
higher dimensions, $\pv$, follows the same line of thoughts.  Two
essential particularities we retain from the one-dimensional
demonstration, because they are important for the thermodynamical
treatment: a) the coefficient matrices $\bf G$ and $\bf D$ can depend
on the energy functional $E(\pv)$ given independently from the
stochastic dynamics, and b) the damping force is proportional to the
velocity in phase space, $\vv=\nabv E$.  In order to treat a quite
general situation, we shall allow additional $\pv$ dependence of the
coefficients, that is, $\pv$ enters not only through the
energy.  Furthermore, we include forces other than damping
proportional to the group velocity, $-{\bf G}\nabv E $, namely,
conservative forces given by ${\bf S}\nabv E $.

\vs The subsequent theory should be compared to the recent works of Ao
\cite{Ao03, Ao04}, where in a sense the opposite path is traveled.
There a general Langevin equation was started with, assuming a unique
stationary distribution.  Taking the latter to be BG, i.e., the
exponential of a potential like in (\ref{FORMAL-GIBBS}), it was shown
that there was an equivalent equation of motion containing the
gradient of that potential, with the usual fluctuation-dissipation
relation holding between diffusion and dissipative mobility
coefficients.  In what follows we depart from the BG recipe, show how
to treat the potential, i.e. the energy, without the restriction to
the BG form, and present the ensuing generalized
fluctuation-dissipation theorem, encompassing also non-BG-type
stationary distributions. 
  
\vs 
The generalized Langevin equation has now the form 
\be
\dot{\pv} + {\bf G}(E,\pv) \nabv E -  {\bf S} \nabv E 
 - \av(E,\pv)= \xiv.
\label{GEN-LANG}
\ee 
We assume that the noise term is Gaussian, white, with 
\bse \ba \langle
\xi_i(t) \rangle &=& 0, \\ \langle \xi_i(t)\xi_j(t') \rangle &=& 2
D_{ij}(E,\pv)\,\delta(t-t'),
\ea \label{NOISE-PROPERTIES} \ese
where $\bf D$ is obviously symmetric. Furthermore, the damping
coefficient ${\bf G}$ is a symmetric matrix, $\av$ is arbitrary apart
from the fact that it is not proportional to $\nabv E$, and the
Hamiltonian, i.e., reversible, forces are characterized by the
antisymmetric, constant matrix, ${\bf S}$.  If the state vector $\pv$
contains canonical conjugate momenta and coordinates then in that
subspace $\bf S$ is the antisymmetric block unit (symplectic) matrix,
while purely relaxational variables are those in whose subspace $\bf
S$ vanishes.  Again, we understand the Langevin equation in the Ito
sense.  We call the reader's attention to the double role of the
energy function $E(\pv)$, namely, it serves as the Hamiltonian
generating the conservative forces ${\bf S} \nabv E$ and, at the same
time, its gradient acts as velocity stream in phase space and so it
participates in linear damping.  The latter encompasses the case of
conventional damping proportional to the group velocity, because then
the damping matrix $\bf G$ is nonzero in the subspace of physical
momenta but vanishes on the canonical coordinates.

\vs The associated Fokker-Planck equation can be written in the form
of a continuity equation,

\be \partial_t f + \nabv \Jv = 0,
\label{CONTINUITY}
\ee 

where the current $\Jv$ is given by

\bse \label{CURRENT-ALL}\ba \Jv &=& \Jv_c +\Jv_d,
\label{CURRENT} \\
\Jv_c &=&  ({\bf S} \nabv E) f(\pv,t),\label{CURRENTC} \\
\Jv_d &=& \left[ \av(E,\pv)- {\bf G}(E,\pv) \nabv E \right] f(\pv,t) -
\nabv \left[{\bf D}(E,\pv) f(\pv,t)\right].
\label{CURRENTD}  \ea \ese 
  
Here we decomposed the current into a conservative and a dissipative
part.  This gains significance when one determines the stationary
distribution $f_s(\pv)$, about which we again make the crucial
assumption $f_s(\pv)=f(E(\pv))$.  Then the stationary current should
satisfy $\nabv \Jv=0$.  The principle of detailed balance
\cite{vanKampen} obviously cannot be invoked here as a condition for
the vanishing of the full current, because the conservative current
cannot be generally suppressed in the presence of Hamiltonian forces,
$\Jv_c\ne {\bf 0}$, as it represents reversible motion within an
energy shell.  It is, however, consistent to assume that the
dissipative part of the current vanishes $\Jv_d={\bf 0}$.  Then again
a purely $E$-dependent stationary distribution $f(E)$ is sought.
Regarding the $E$-dependence, detailed balance means that the
coefficient matrix of the $\nabv E$ term and the rest in
(\ref{CURRENTD}) have to vanish separately as
  
\bse\label{FLUCT-DISS-ALL}\ba {\bf G}(E,\pv) &=& {\bf D}(E,\pv)/T_{\slp}(E) -
\partial_E {\bf D} (E,\pv),
\label{FLUCT-DISS}\\
\av(E,\pv) &=& \nabv ' {\bf D}(E,\pv), \label{FLUCT-DISS2} 
\ea \ese

where $\nabv '$ means gradient in terms of the explicit $\pv$
dependence, and $T_{\slp}(E)$ characterizes the stationary
distribution as defined in eq.\ (\ref{INV-LOG-SLOPE}). Apparently
(\ref{eq:FDT-1d}) is the special case of (\ref{FLUCT-DISS}).  The
conservative current, $\Jv_c$, does not affect the equilibrium energy
distribution, because at stationarity, that is when
$f(\pv,t)=f_s(\pv)=f(E(\pv))$, the condition $\nabv \Jv_c = {\bf 0}$ is
automatically fulfilled due to the antisymmetry and constancy of the
matrix ${\bf S}$.  It is far from trivial that between the matrices
$\bf G$ and $\bf D$, which are in principle calculable from a
microscopical dynamics, the relation (\ref{FLUCT-DISS}) would contain
an energy dependent scalar factor only.  This strict requirement,
together with eq.~(\ref{FLUCT-DISS2}) on the nongradient damping
$\av$, is the fluctuation dissipation theorem for the generalized
thermodynamical equilibrium. We can also ascertain that the existence
of the nonlinear damping force $\av$ is due to the extra
$\pv$-dependence of the diffusion matrix; if the latter depends on the
state vector $\pv$ only through the energy then the stationary
distribution of the form $f(E)$ excludes such a force.
  
\vs The fluctuation-dissipation theorem (\ref{FLUCT-DISS}) can be used
to obtain the diffusion matrix $\bf D$ once $\bf G$ is known, like it
was done in the one-dimensional case in the previous section.  The
analog of (\ref{D-FROM-G-1d}) now is

\be {\bf D}(E,\pv) = \frac{1}{f(E)} \int_E^{\infty} {\bf
  G}(\bar{E},\pv) f(\bar{E}) d\bar{E}.
\label{D-FROM-G}
\ee 

If the damping matrix does not depend on the state variable through
the energy then 

\be {\bf D}(E,\pv) =  T_{\mathrm E}(E) {\bf
  G}(\pv), \label{EinsteinRelation-anyd}
\ee 

wherein the Einstein temperature (\ref{EinsteinTemperature}) enters.

\vs The BG distribution corresponds to the $T_{\mathrm E}(E)= T_{\slp}(E)=T$
constant case.  This can be achieved, in principle, also by
energy-dependent damping and diffusion coefficient matrices; so the
fact that these coefficients depend on the energy of the observed
mesoscopic slow mode does not necessarily imply that the BG
distribution is not applicable. On the contrary, whether this is the
case, should be decided on the basis of the underlying microscopical
dynamics. In many cases it is relatively easy to obtain the damping
coefficient, but hard to calculate the diffusion and other noise
properties.  The Tsallis distribution can be obtained, similarly to
the one-dimensional case, by a constant $\bf G$ and a linear ${\bf
  D}(E)$, but as a quite strong condition for all, only $E$-dependent,
stationary distributions also the corresponding matrices have to be
proportional.

\vs
\section{Hadron spectra in heavy ion collisions} 
 
\subsection{Data fits} 
 
\vs We now turn to extracting the energy distribution and its main
parameters, with special attention to $T_{\slp}(E)$, from experimental data.
In most cases of observed thermal systems the BG distribution has been
recovered, and it is seldom expected to find something else.  On the
other hand a power-law can be well-fitted to experimental heavy-ion
high transverse momentum yields of neutral pions.    The data available
for the highest energy are for neutral pions from the PHENIX
collaboration \cite{PTLexperiments},  so we mainly focus on these
spectra. 
 
\begin{figure}  
\begin{center}  
\includegraphics[width=0.65\textwidth,angle=0]{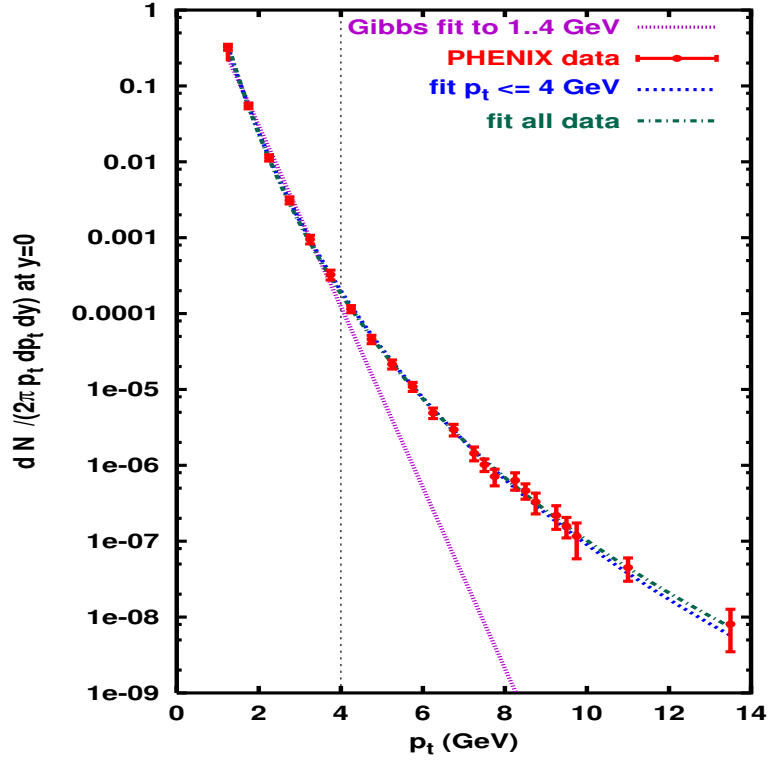}  
\end{center}  
\caption{\label{Figminpionpt}  
  The minimum bias neutral pion $p_{\tr}$ spectrum from Au-Au
  collisions at 200 GeV/nucleon from the PHENIX group. Displayed is
  the spectrum $\dt N /(2\pi p_{\tr} \dt p_{\tr} \dt y)$ at $y=0$ as
  function of the transverse momentum $p_{\tr}$.  Experimental data
  (large dots with solid error bars) are used in the range 1-4 GeV for
  fit to BG (dotted straight line) and for Tsallis (dashed curve)
  distributions, as well as a Tsallis fit over all data points
  (dash-dotted curve), the latter having $T= 118$ MeV, $q=1.1173$.
  The BG fit is manifestly inadequate, while the two Tsallis functions
  remain quite close to each other, so the fit in 1-4 GeV yields a
  surprisingly good prediction for higher energies. }
\end{figure}

\vs Figure \ref{Figminpionpt} presents minimum bias neutral pion data
on the transverse momentum spectrum as it was obtained by the PHENIX
group in RHIC experiment in the mid-rapidity window ($y\approx 0$) by
filled circles.  We test the assumption that this would reflect a
stationary distribution, which only depends on the energy  

\be
\frac{\dt N^{(m)}}{2\pi p_{\tr} \dt p_{\tr} \dt y} = A \: f(E(p_{\tr})),
\label{NoflowSpectra}
\ee
  
with a normalization constant $A$.  As a forward reference, this
follows from eq.~(\ref{integral}) without transverse flow, $v=0$.  For
free relativistic particles with rapidity $y$ the energy is given by
$E(p_{\tr})=\cosh(y)\sqrt{(m^2 + p_{\tr}^2)}$. In the aforementioned
rapidity window the average $\langle\cosh^2y\rangle = (1+\sinh(1))/2
\approx 1.087$ has only a negligible effect.  The BG distribution is
fitted in the low-momentum range of $1$ -- $4$ GeV resulting in an
inverse slope of $T = 364$ MeV.  This high temperature is
conventionally interpreted as a consequence of a blue shift factor of
nearly $2$ due to transverse flow.  Tsallis distribution fits are done
both in the above low $p_{\tr}$ range and for all data: these two
curves are quite close to each other. This means that the Tsallis fit
reveals a remarkable predictive power for these data.  The parameters
of the full fit were $T= 118$ MeV, $q=1.1173$.  While this $T$ is
significantly lower than the color deconfinement transition range for
subcritical chemical potentials $162-164$ MeV (see \cite{Fodor04} and
refs.\ therein), we emphasize that the two kinds of temperature cannot
be directly compared, because we fitted the Tsallis parameter, whereas
the transition temperature was computed from BG statistics.  We
speculate that the assumption of an a priori BG ensemble may have to
be released in future lattice calculations. 
  
\begin{figure}
\begin{center}
\includegraphics[width=0.5\textwidth,angle=-90]{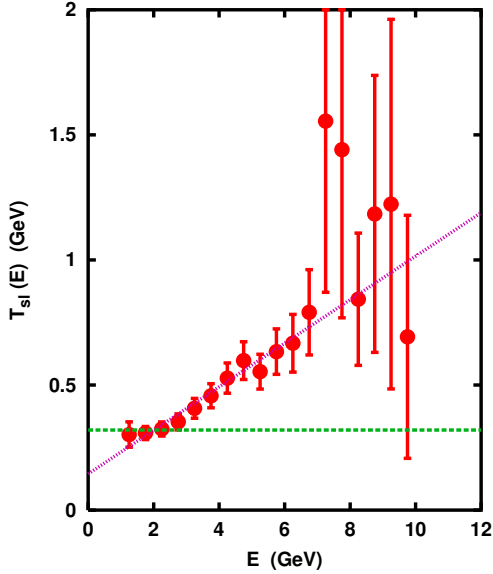}
\end{center}
\caption{\label{Figpi0slopemin}
  The inverse slope of the neutral pion spectrum for minimum bias as
  function of $E=m_{\tr}$.  The data (points with solid error bars)
  are obtained by numerical derivation from the data in Fig.\ 
  \ref{Figminpionpt}, the BG constant at $320$ MeV is a guide to the
  eye for the low-energy part (horizontal line), and the linear,
  increasing, function is the Tsallis fit, with $T=165$ MeV and
  $q=1.094$.}
\end{figure} 
  
 
\vs In Fig.~\ref{Figpi0slopemin} we plot the spectral temperature,
i.e., the inverse logarithmic slope $T_{\slp}(E)$, as defined in
eq.~(\ref{INV-LOG-SLOPE}), obtained from the neutral pion cross section
of Fig.\ \ref{Figminpionpt} by numerical derivation versus the
transverse mass $E=m_t=\sqrt{(m^2 + p_{\tr}^2)}$.  This choice of the
energy variable $E$ does not take into account transverse flow
corrections.  The error bars in $T_{\slp}(E)$ were computed from the
statistical error in $\dt N/(2\pi p_{\tr}\dt p_{\tr} \dt y)$ 
by assuming Gaussian error propagation.  For these data the
constant $T_{\slp}(E)=T$, that is, the BG distribution, is definitely
excluded.  The transformed data points do show a monotonous rise, but
with a distinct curvature in the low-energy part.  Since this is in
the flow-sensitive region, and presently flow effects are neglected,
it would not make sense to try to reproduce the curve.  We give a
linear, i.e., Tsallis, fit only for the range $1-7$ GeV, because for
higher energies the error is becoming large.  The Tsallis fit is
compatible with the data points considering the error bars, but the
uncertainty of the fit is demonstrated by the difference in the fitted
value of the Tsallis temperature $T\approx 165$ MeV from that in 
Fig.~\ref{Figminpionpt}.  The present fitting is obviously not deciding
about the precise shape of $T_{\slp}(E)$, however, it shows in
tendency a predominantly linear deviation from the assumption of a
constant, i.e., BG temperature function.
  
\vs We tested the consistency of the above fit by separately
considering different centrality bins: all $\pi^0$ data seem to have
been fitted by one $T_{\slp}(E)$ curve irrespective of centrality.
Therefore we think it made sense to use minimum bias data for a common
fit, the more so because that way the statistical errors were better
suppressed.

\subsection{Flow corrections} 
     
\vs Given the importance of the low momentum range for the overall
fit, we also have to take into account the transverse flow.  First we
recapitulate the standard treatment of a collective flow emitting
hadrons \cite{CsernaiBook}.  We shall consider differential spectra of
hadrons, identified via the rest mass $m$, detected with a certain
transverse momentum, $p_{\tr}$, azimuthal angle $\varphi$ and longitudinal
rapidity $y$.  The four-momentum of such a hadron is given by \be p =
(m_{\tr}\cosh y, \: m_{\tr} \sinh y, \: p_{\tr} \cos\varphi, \:
p_{\tr}\sin\varphi), 
\label{pdef}
\ee 
with $m_{\tr}=\sqrt{m^2+p_{\tr}^2}$. Such hadrons may come from different
space-time points of an emitting source. The position of the emission
is characterized by the four-vector,
\be
 x = (\tau\cosh\eta, \: \tau\sinh\eta, \: r \cos\psi, \: r\sin\psi).
\label{xdef}
\ee
The total number of detections is described by an eight-dimensional
integral of the one-particle phase space density, essentially
the Wigner function: 
\be 
N \, = \, \int\! \dt^4x  \int\! \dt^4p \, \quad  W(x,p).
\label{Wigner} 
\ee  
We adopt here the widespread assumption that the emission of the 
finally detected hadrons happens at a certain longitudinal proper 
time, the break-up time $\tau=\tau_{\mathrm{br}}$, but otherwise  
homogeneously in a cylindrical three-volume, $V=\pi R^2L$, with 
radius $R$ and length $L=2\tau_{\mathrm{br}}\eta_0$. Experimental data 
are available for transverse spectra of identified hadrons, i.e. for 
\be
 \frac{\dt N^{(m)}}{p_{\tr}\dt p_{\tr} \dt\varphi \dt y} \, = \,
 mV \int_{-\eta_0}^{\eta_0} \frac{\dt\eta}{2\eta_0} 
 \int_{-\pi}^{\pi} \frac{\dt\psi}{2\pi} \, f(E),
\label{integral}
\ee
where the energy distribution $f(E)$ is the thermodynamical
distribution we are seeking to learn about. The energy-variable, $E$,
in this distribution is the energy in the local Lorentz frame,
$E=u\cdot p$, with $u$ being the four-velocity of the
emitting cell. (In the case $f(E)$ is the BG distribution, this version
is called the J\"uttner distribution.)

\vs A further simplifying assumption is to consider a scaling
longitudinal flow (the so called Bj\o rken flow) combined with a
transverse radial expansion of velocity $v$. The four-velocity
becomes
\be u \, = \, (\gamma\cosh\eta, \: \gamma\sinh\eta, \: \gamma\,
v\cos\varphi, \: \gamma \,v\sin\varphi)
\label{udef}
\ee
with $\gamma=1/\sqrt{1-v^2}$. In this case
\be E \, = \, \gamma\, m_{\tr} \cosh(y-\eta) \: - \: \gamma\, v\,p_{\tr}
\cos(\varphi-\psi).
\label{Eflow}
\ee
According to eq.~(\ref{integral}) the detectors see an average of the
local thermodynamical distribution over $\eta$ and $\psi$.  This
integral can be calculated analytically for the BG distribution
resulting in a product of the $K_0(\gamma m_{\tr})$ and $I_0(\gamma v
p_{\tr})$ Bessel functions, but it is not so in the Tsallis or in more
general cases.  Numerical integration is doable by using suitable
ans\"atze for $f(E)$, but then one looses the possibility of direct
reconstruction of this function from observational data.  Another way
is to make yet further simplifying assumptions on the emitting source,
but keeping a general $f(E)$.

\vs Following the second reasoning we adopt the physical picture that
the main contribution to the detected spectrum is due to forward
emission, that is, it comes from those particles that fly in the same
direction as the emitting volume element.  Thus we replace the $E$
function by its value at $\eta=y$ and $\psi=\varphi$, i.e.,
$E=\gamma(m_{\tr}-vp_{\tr})$.  We learn that the shape of the observed
transverse distributions at large $p_{\tr}$ is effected only by an overall
blue-shift, $p_{\tr}=E\sqrt{\frac{1+v}{1-v}}$ , while at small $p_{\tr}$
mass-dependent effects occur.  For instance, in the non-relativistic
limit $f(E(p))$ is Gaussian in $p_{\tr}$ such that the variance depends on
the rest mass as $\langle p_{\tr}^2 \rangle = mT^\ast = m(T_0 + mv^2)$,
where $T^\ast$ is the empirical temperature and $T_0=T_{\slp}(E=m)$ is the
local inverse slope at rest.  This picture is in general accordance
with the widely accepted role of the hydrodynamical flow on the
particle spectrum.
 
\begin{figure} 
\begin{center} 
\includegraphics[width=0.5\textwidth,angle=-90]{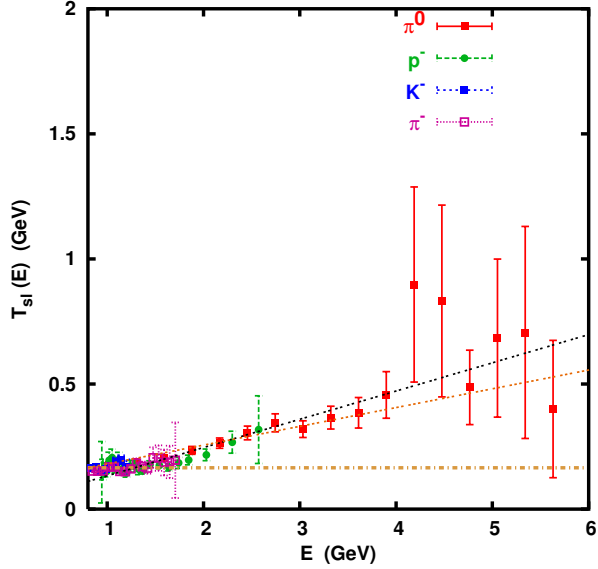}
\end{center} 
\caption{\label{Figallflow}
  Inverse logarithmic slopes obtained from transverse flow corrected
  $p_{\tr}$ spectra of various particles.  The flow corrected energy
  $E = \gamma(m_\tr - vp_\tr)$ is on the horizontal axis, while
  $T_\slp(E) = \dt E/\dt m_\tr \, T_\slp(m_\tr) = \gamma(1 - v
  m_\tr/p_\tr)\, T_\slp(m_\tr)$ on the vertical one, where
  $T_\slp(m_t)=-f/(\dt f/\dt m_\tr)$, and $f(m_t)$ were taken from the
  experiments.  The particle data $\pi^0$ (squares with solid error
  bars), antiprotons (large dots with dashed bars), $K^-$ (squares
  with short dashed bars), $\pi^-$ (open squares with dotted bars)
  were fitted with a common $v=0.5$ to two Tsallis linear functions,
  (1) with $T=20.5$ MeV and $q=1.26$ (short dashed lines grouped in
  two) and (2) with $T=113$ MeV and $q=1.081$ (short dashed lines
  grouped in three).  The BG line (dash-dotted) is horizontal,
  obtained from a fit for $E<1.5$ GeV. }
\end{figure}

\vs Fig.~\ref{Figallflow} shows the inverse logarithmic slope,
$T_{\slp}(E)$ extracted from antiproton, negative kaon and pion, and
neutral pion data from \cite{PTLexperiments} with the assumption
$E=\gamma(m_{\tr}-vp_{\tr})$.  The curves for different hadron types
fit so well together with a common $v=0.5$ that this strongly supports
the assumption about the statistical origin of the local $f(E)$
distribution.  A faster transverse flow, e.g. $v=0.6$ already
overcompensates the low-$p_{\tr}$ data for antiprotons turning the
temperature $T_{\slp}(E)$ abruptly down towards negative values,
therefore we stick with the maximal flow value of $v=0.5$ not showing
the downturn. The Tsallis fits, which may be layed on the data are not
fine enough; a surprisingly low temperature parameter and a relatively
high deviation from the BG distribution with $T=20.5$ MeV and $q=1.26$
(Tsallis fit 1) goes through the data points with error bars as well
as the Tsallis fit 2 with $T=113$ MeV and $q=1.081$. In the latter
case at the lowest $p_\tr$ data the virtual slope is $T_{\slp}(0.825
GeV) = 165.5$ MeV, roughly equal to the conventional BG estimate.
 
 
\vs It is clear, however, that more work is needed to go beyond our
crude estimate of the flow's effect, possibly in an iterative manner
with ever better approximations for the stationary distribution.  On
the other hand, since higher $p_{\tr}$ data at RHIC are expected, we are
looking forward to learn more about a possibly non-BG behavior in a
momentum range where transverse flow effects little 
distort the local slope in particle spectra.  

\section{Conclusion}  \vs In this paper we considered
generalized stationary energy distributions in the case of energy
dependent damping and noise.  We have shown that the BG and Tsallis
distribution are the most simple particular cases of this framework.
As a physical example, high energy transverse momentum distributions
have been analyzed from relativistic heavy ion experiments at RHIC.
The high energy part of the neutral pion data (above $p_{\tr} \approx 2 -
4$ GeV) excludes the Boltzmann-Gibbs distribution.  This fact earlier
has been interpreted as a sign for a non-equilibrium component in the
spectra \cite{Fries03a, Fries03b}.  Recent field theoretical model
estimates on the speed of thermalization, however, predict at least
partial equilibration, called ``prethermalization'' \cite{Patkos04}.
According to our ideas presented in this paper, a generalized
equilibrium interpretation of these spectra becomes thinkable.
Finally we note that the best fit to the data, while definitely
excludes the BG distribution, at this stage does not decisively
support the Tsallis distribution either.  A more general distribution
based on a quadratic energy dependence of the diffusion coefficient is
also a possible interpretation. In this respect further investigations
are needed to clarify the deformation effect of transverse flow on the
discussed spectra.    Higher particle yields are also desirable
because they can lead to narrowing error bars for larger energies.  

\vs Finally we stress that our empirical findings are reminiscent to the
result reported about by Walton and Rafelski \cite{RafelskiWalton} on
charmed quark emission, where a predominantly linear $T_{\slp}(E)$ has
been obtained from perturbative QCD, albeit with a slight curvature. 
This raises the hope for a theory that describes microscopically the
quark-gluon-plasma, produces a foundation for the Fokker-Planck
approach and connects it to the detected spectra.

\begin{acknowledgments} 
\vs Discussions with Professors J. Zim\'anyi and L. Sasv\'ari are
gratefully acknowledged.  This work has been supported by the
Hungarian National Research Fund OTKA (T 034269 and TS 044839)
and by the Deutsche Forschungsgemeinschaft through a Mercator
Professorship for T.S.B.
\end{acknowledgments}


\bibliography{ngps} 
\bibliographystyle{apsrev}

\end{document}